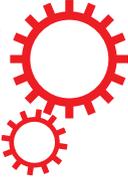

# Potential of mean force between like-charged nanoparticles: Many-body effect

Xi Zhang[1,*], Jin-Si Zhang[1,*], Ya-Zhou Shi[1], Xiao-Long Zhu[2] & Zhi-Jie Tan[1]

Ion-mediated interaction is important for the properties of polyelectrolytes such as colloids and nucleic acids. The effective pair interactions between two polyelectrolytes have been investigated extensively, but the many-body effect for multiple polyelectrolytes still remains elusive. In this work, the many-body effect in potential of mean force (PMF) between like-charged nanoparticles in various salt solutions has been comprehensively examined by Monte Carlo simulation and the nonlinear Poisson-Boltzmann theory. Our calculations show that, at high 1:1 salt, the PMF is weakly repulsive and appears additive, while at low 1:1 salt, the additive assumption overestimates the repulsive many-body PMF. At low 2:2 salt, the pair PMF appears weakly repulsive while the many-body PMF can become attractive. In contrast, at high 2:2 salt, the pair PMF is apparently attractive while the many-body effect can cause a weaker attractive PMF than that from the additive assumption. Our microscopic analyses suggest that the elusive many-body effect is attributed to ion-binding which is sensitive to ion concentration, ion valence, number of nanoparticles and charges on nanoparticles.

Ions play critical roles in the structure and stability of charged systems such as colloids and nucleic acids[1–3]. Due to like-charged nature, the aggregation of colloids or the structural folding of nucleic acids would incur strong Coulombic repulsion, while salt ions in solutions can reduce such like-charged repulsion, and even drive the conformation to compact structures. Therefore, the ion-mediated effective interaction is essential to the energetics of colloid stabilization and nucleic acid structural folding[4–9].

As a typical paradigm, the systems of two like-charged polyelectrolytes such as two colloids and two DNAs in ionic solutions have attracted considerable interests and have been studied extensively in recent years[10–17]. Previous studies have suggested that monovalent ions generally modulate an effective repulsion between two like-charged particles[5,6,18–21] and may possibly induce aggregation of like-charged rods at some special conditions with strong electrostatic correlations[22]. However, multivalent ions can generally cause an effective attraction[5,11,23,24], and such multivalent ion-mediated attraction is attributed to the strong charge correlation of multivalent ions and polyelectrolytes, and has been proposed to be the driving force for DNA condensation and RNA structural collapse[25–28]. Nevertheless, realistic systems are generally composed of many like-charged particles such as colloids and DNA helices[29–32]. Is the pair effective interaction between two like-charged particles additive for many-body particle system?

To understand such many-body effect, some investigations have been performed for rodlike and spherical polyelectrolytes in ion solutions[18,20–22,33–36]. Some experiments with optical tweezers have been employed to accurately control three-body macroion systems in monovalent ion solutions, in which the pair interactions are found to be repulsive and more macroions can reduce such repulsion in dilute salt[18,20,21]. Meanwhile, some classic polyelectrolyte theories have been employed on the many-body effect. The counterion condensation (CC) theory has been extended for many-body like-charged rods to show the non-additivity of the potential of mean force (PMF) between charged rods, while the theory always predicts an effective attraction between rods even at low monovalent salt[37]. The Poisson-Boltzmann (PB) theory[38–41] has been widely employed for charged systems and its linearized version has also been used to probe the many-body effect for colloidal particles and for rodlike polyelectrolytes in monovalent salt solutions[18,20,21,33], and the predicted repulsive PMF appears non-additive at low monovalent salt[18]. However, due to the neglect of inter-ion correlations and the linearization of nonlinear

[1]Center for Theoretical Physics and Key Laboratory of Artificial Micro & Nano-structures of Ministry of Education, School of Physics and Technology, Wuhan University, Wuhan 430072, China. [2]Department of Physics, School of Physics & Information Engineering, Jianghan University, Wuhan 430056, China. *These authors contributed equally to this work. Correspondence and requests for materials should be addressed to Z.J.T. (email: zjtan@whu.edu.cn)

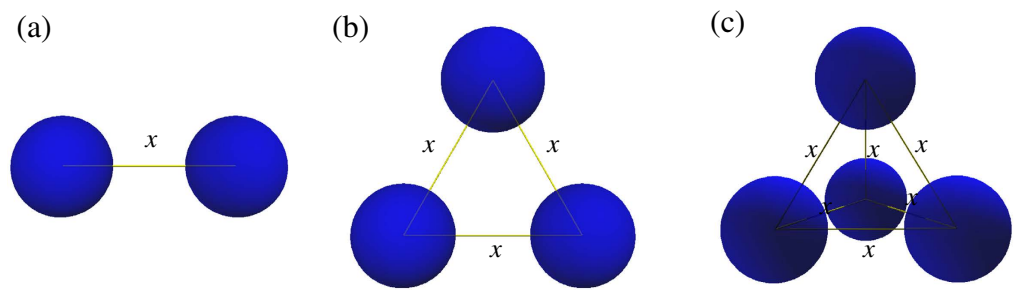

**Figure 1.** (**a**) Two-body system with two nanoparticles in a line; (**b**) Three-body system with three nanoparticles in the configuration of equilateral triangle; (**c**) Four-body system with four nanoparticles in the configuration of equilateral tetrahedron. In each configuration, the nanoparticles are completely spatially equivalent. Such configurations are suitable for analyzing the many-body effects in potential of mean force between like-charged nanoparticles.

Boltzmann term, the linearized PB theory cannot properly treat multivalent ion-polyelectrolyte interaction and the highly charged polyelectrolytes with strong electric field nearby. Nevertheless, many polyelectrolytes are generally highly charged and multivalent ions can be very important for strong polyelectrolyte systems, such as nucleic acids and F-actins[4,10,42,43].

Although some advanced theories have been proposed such as the dressed ion theory and strong coupling theory for charged colloids[44,45], the integral equation theory for polyelectrolyte[46], and the tightly bound ion (TBI) theory for nucleic acids[47], little attention has been paid to the many-body effect for polyelectrolytes in multivalent salt solutions so far. The charge-fluctuation theory predicts that the attractive many-body PMF at low divalent salt would cause polyelectrolyte aggregation with no size limit while high divalent salt cannot induce polyelectrolyte aggregation[48], which is somewhat in contradiction with the experiments on DNA/F-actin condensation[42,49]. The TBI theory has been employed for triple-DNA system to explore the free energy landscape for helix assembly, but the many-body effect is still required to be examined[24]. As a complementary bridge between theories and experiments, computer simulations have become powerful tools to probe the effective interaction between nanoparticles and between nucleic acids in solutions[5,6,13,14,16,22,23,32,34–36,50]. The Monte Carlo (MC) simulation has been applied to the system of three-body colloids in monovalent salt and it is shown that the additive assumption overestimates the repulsive PMF between charged colloids at monovalent salt lower than 0.05M[19], while a coarse-grained molecular dynamics simulation with charged dendrimers shows that many-body effect can enhance the pair repulsion in monovalent ion solutions[34]. Furthermore, the many-body effect for polyelectrolytes in multivalent salt solutions has been seldom covered. Therefore, until now, there is still lack of an overall picture on the many-body effect in the ion-mediated interaction between like-charged particles over wide ranges of ionic conditions, especially in multivalent salt solutions.

In this work, we will investigate the many-body effect in the PMF between like-charged nanoparticles by the MC simulation with the thermodynamics-integration (TI) method, as well as the nonlinear PB theory. Beyond the previous studies[18–21,34–36], the present work will cover wide ionic conditions and focus on the multivalent salt solutions which have been rarely paid attention to, in order to provide a comprehensive understanding on the many-body PMF between like-charged polyelectrolytes.

### Results and Discussions

In this work, we will calculate the many-body PMFs $\Delta G_x$ by MC simulation and the nonlinear PB theory for two-body, three-body and four-body like-charged nanoparticles in 1:1 and 2:2 salt solutions, as displayed in Fig. 1. We will emphasize the many-body effect on the PMFs over the wide ionic conditions of realistic interests and the corresponding microscopic mechanism. In the following, four-body systems are considered as the reference systems for analyzing the many-body effect.

**Many-body PMFs in 1:1 salt.** *PMFs are non-additive at low 1:1 salt and additive at high 1:1 salt.* The MC simulation with TI method used in this work has been extensively tested and validated for two like-charged nanoparticles in 1:1, 2:2 and 3:3 salt solutions; see Figure S1 in Supplementary Information (SI). For 1:1 salt, as shown in Fig. 2a–c, the PMFs between like-charged nanoparticles are always repulsive, and appear strongly dependent on ion concentration. At low (~0.005M) 1:1 salt, the repulsive PMF is very strong, while at high (~0.5M) 1:1 salt, the repulsive PMF become much weaker. The many-body effect is also strongly dependent on 1:1 salt concentration. As shown in Fig. 2a, the additive assumption with two-body PMF would apparently overestimate the repulsive PMF for four-body nanoparticles at low (~0.005M) 1:1 salt, i.e., the many-body effect would weaken the pair repulsion between two nanoparticles. Specifically, such overestimation can reach as high as ~45% at center-to-center separation $x = 25$ Å, and is expected to become more pronounced at lower 1:1 salt (than 0.005M 1:1 salt). As salt concentration is increased to ~0.05M, the many-body effect of PMF becomes relatively weaker, and the additive assumption with two-body PMF overestimates the repulsive PMF of four-body system at $x = 25$ Å by ~26%; see Fig. 2b. When salt concentration becomes sufficiently high (~0.5M), the PMF between nanoparticles becomes approximately additive and the many-body effect becomes negligible, as shown in Fig. 2c.

To understand such many-body effect, we would like to analyze ion-binding through calculating the net ion charge fraction $Q(r)$ within a distance $r$ from the centers of nanoparticles

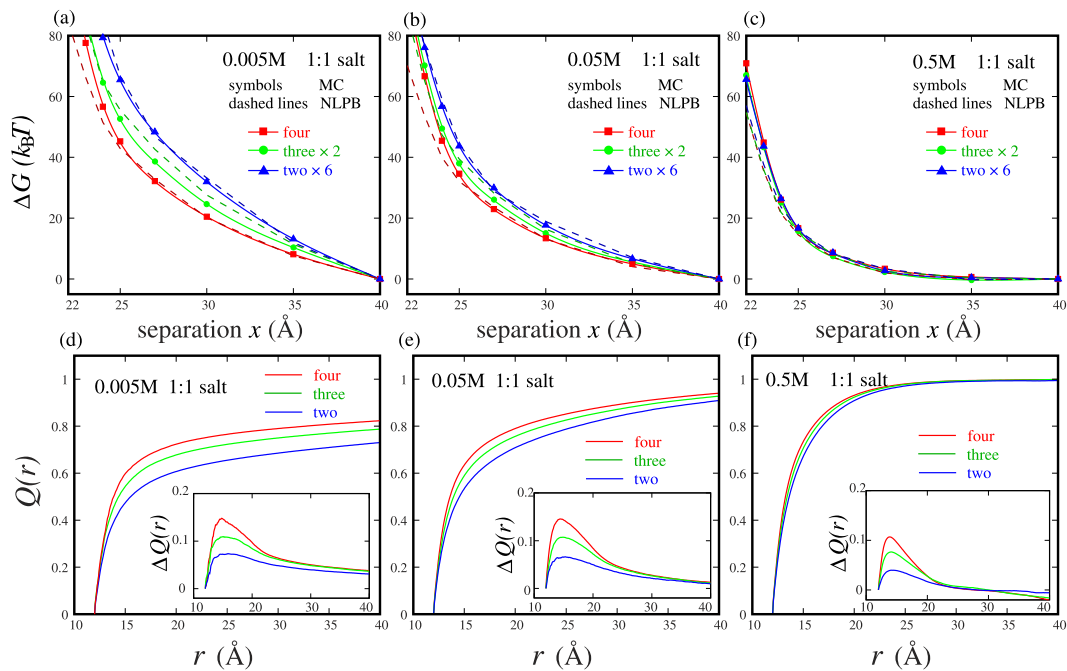

**Figure 2.** (**a**–**c**) The potentials of mean force $\Delta G_x$ as functions of the separation $x$ between nanoparticles with $-24e$ for the four-body systems in 1:1 salt solutions which are calculated respectively by the pair-wise potential of mean force abstracted from two-body, three-body and four-body systems (denoted respectively by two nanoparticles $\times 6$, three nanoparticles $\times 2$, and four nanoparticles). (**a**) 0.005M, (**b**) 0.05M, and (**c**) 0.5M. The PMFs from the nonlinear PB are shown as comparisons (dashed lines). (**d**–**f**) Net charge distribution $Q(r)$ per unit charge on nanoparticles as a function of distance $r$ around the nanoparticles with $x = 25$ Å in 0.005M (**d**), 0.05M (**e**) and 0.5M (**f**) 1:1 salt solutions. The insets show the increase of $Q(r)$ due to the approaching of nanoparticles from $x = 40$ Å to $x = 25$ Å for the systems of two nanoparticles, three nanoparticles and four nanoparticles.

$$Q(r) = \frac{1}{M|Z|} \int_{<r} \sum_\alpha z_\alpha c_\alpha(\mathbf{r}) \mathrm{d}^3 \mathbf{r}, \tag{1}$$

where $M$ is the number of nanoparticles. $Z$ is the charge on nanoparticles. $z_\alpha$ is the valence of $\alpha$ ion species and $c_\alpha(\mathbf{r})$ denotes its concentration on position $\mathbf{r}$ of Cartesian coordinate system.

We use $Q(r)$ instead of the counterion condensation values of the CC theory, since $Q(r)$ can describe the radial distribution of ion neutralization fraction and the ions in the vicinity of nanoparticle surface would make the major contribution to ion-nanoparticle interactions[23,47,51]. As shown in Fig. 2d–f, ion-binding depends strongly on ion concentration. At low 1:1 salt, a charged nanoparticle only gets weak ionic neutralization due to the high entropy penalty for ion-binding. At 0.005M 1:1 salt, two-body nanoparticles with center-to-center separation of $x = 25$ Å can only get ~57% ionic neutralization by $1+$ ions within 8 Å (two layers of ions) from nanoparticle surface. Thus, the pair PMF between two nanoparticles is strongly repulsive due to the weak ionic neutralization. But for many-body nanoparticles, the involvement of more nanoparticles would greatly enhance the electric field, especially in the region between them. At 0.005M 1:1 salt with $x = 25$ Å, the increase in neutralization fraction with the number of nanoparticles increased from 2 to 4 is ~12% within 8 Å from nanoparticle surface. More nanoparticles would cause many more binding ions between them and consequently would result in stronger ionic neutralization. As a result, at low 1:1 salt, the many-body PMF is apparently less repulsive than that calculated from the additive assumption with the pair PMF between two-body nanoparticles; see Fig. 2a. As ion concentration is increased (e.g., to ~0.05M), the ion-binding would become stronger due to lower entropy penalty, and two-body system can get stronger ionic neutralization. Consequently, the enhancement of ion-binding by more nanoparticles would become weaker, and thus the many-body effect in PMFs would become less significant for higher 1:1 salt concentration. When ion concentration becomes very high (~0.5M), the nanoparticles can get very strong ionic neutralization. As shown in Fig. 2f, two nanoparticles with $x = 25$ Å can get ~85% ionic neutralization by $1+$ ions within 8 Å from nanoparticle surface and involvement of more nanoparticles can only cause nearly invisible enhancement in ion-binding. Therefore, the many-body PMF is nearly equal to that calculated from the additive assumption with the pair PMF between two-body nanoparticles and the many-body effect becomes negligible. It is not strange that the ion-binding values shown above are dependent on salt concentration. The extended CC theory has shown that, for rod-like polyelectrolyte, the counterion condensation values are independent on ion concentration, while such values would become dependent on ion concentration for spherical polyelectrolyte[51].

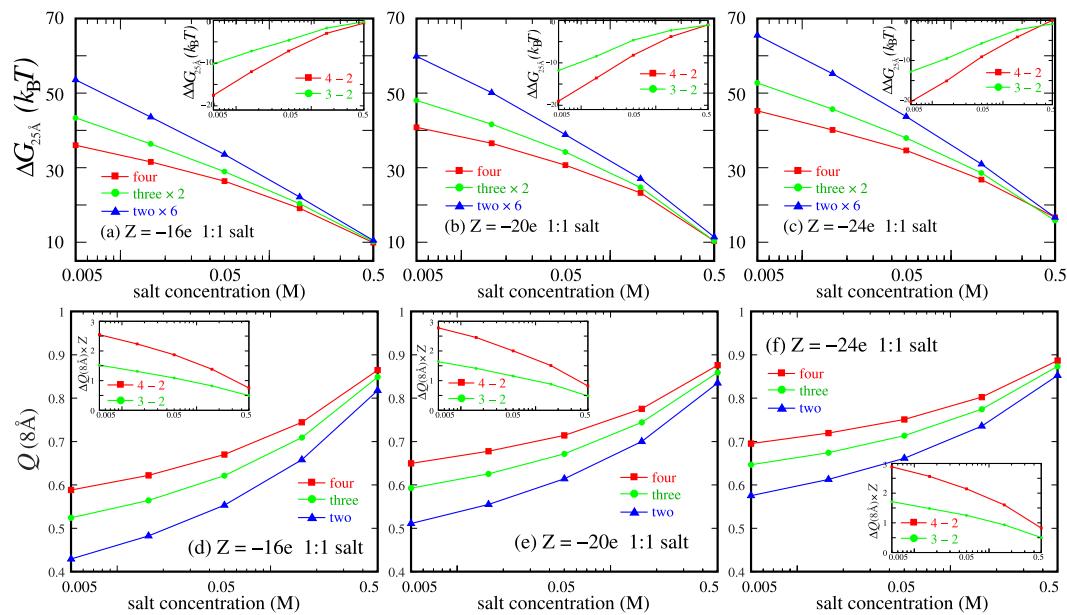

**Figure 3.** (**a**–**c**) $\Delta G_{25\,\text{Å}}$ ($=G_{25\,\text{Å}} - G_{40\,\text{Å}}$) as functions of 1:1 salt concentration for the four-body systems which are calculated respectively by the pair-wise potential of mean force abstracted from two-body, three-body and four-body systems (denoted respectively by two nanoparticles $\times 6$, three nanoparticles $\times 2$, and four nanoparticles). (**a**) $Z = -16e$, (**b**) $Z = -20e$ and (**c**) $Z = -24e$ per nanoparticles. The insets show the PMF differences $\Delta\Delta G_{25\,\text{Å}}$ defined by $\Delta G_{25\,\text{Å}}(\text{four}) - \Delta G_{25\,\text{Å}}(\text{two}) \times 6$ (red; denoted by $4 - 2$) and $\Delta G_{25\,\text{Å}}(\text{three}) \times 2 - \Delta G_{25\,\text{Å}}(\text{two}) \times 6$ (green; denoted by $3 - 2$). (**d**–**f**) Net charge fraction within 8 Å from the surface of nanoparticles with $x = 25$ Å for $Z = -16e$ (**d**), $Z = -20e$ (**e**) and $Z = -24e$ (**f**). The insets show the difference in number of binding ions within 8 Å from the surface of nanoparticles between four- and two-body systems (red; denoted by $4 - 2$) as well as that between three- and two-body systems (green; denoted by $3 - 2$).

The PMFs calculated with the nonlinear PB have also been shown in Fig. 2a–c as comparisons. First, the predictions from the nonlinear PB also indicate that the many-body effect would weaken the pair repulsion at low (~0.005M and ~0.05M) 1:1 salt and such many-body effect becomes insignificant at high (~0.5M) 1:1 salt, which verifies the conclusions from the MC simulations since the nonlinear PB has been considered as a good treatment for 1:1 salt solution due to the relatively weak ion-ion electrostatic correlations involved in the cases[8,11,18,24,47]. Second, the predicted PMFs from the nonlinear PB are very close to those from the MC simulations. For the cases where ion concentration can be extremely high, e.g., the four-body nanoparticles with small separation ($x < 25$ Å) and at high (~0.5M) 1:1 salt, the repulsive PMFs from the nonlinear PB are slightly weaker than those from the MC simulations. This may be because the nonlinear PB ignores the ion exclusion and would overestimate the ion concentration for the strongly correlated cases.

*Many-body effect versus charge density on nanoparticles in 1:1 salt.* To examine how the charges on nanoparticles affect the many-body effect and the generality of the above results, we use different charges of $Z = -16e$ and $-20e$ on each nanoparticle to calculate PMFs for two-body, three-body and four-body systems. Figures 2a–c and S2a–c in SI show that $\Delta G_{25\,\text{Å}}$ ($= G_{25\,\text{Å}} - G_{40\,\text{Å}}$) could nearly determine the strength of repulsive PMFs versus the separation between nanoparticles. As shown in Fig. 3a–c, $\Delta G_{25\,\text{Å}}$ exhibits the similar trend in salt concentration dependence for nanoparticle systems with different $Z$'s: (*i*) at low (~0.005M) 1:1 salt, the many-body effect would weaken the pair repulsion between two nanoparticles and such effect becomes weaker as salt concentration is increased; (*ii*) $\Delta G_{25\,\text{Å}}$'s for two-body, three-body and four-body system would converge at high (~0.5M) 1:1 salt, i.e., the many-body effect becomes negligible for nanoparticles with different charges at high 1:1 salt.

The degree of many-body effect for PMFs can be approximately quantified described by $\Delta\Delta G_{25\,\text{Å}} = \Delta G_{25\,\text{Å}}(\text{four-body}) - \Delta G_{25\,\text{Å}}(\text{two-body}) \times 6$. As shown in the insets of Fig. 3a–c, the global many-body effect becomes stronger for higher charge $|Z|$ on nanoparticles at low salt. For example, at 0.005M 1:1 salt, $\Delta\Delta G_{25\,\text{Å}}$ is ~17.5$k_BT$ for $Z = -16e$ and increases to ~20.3$k_BT$ for $Z = -24e$. As 1:1 salt is increased to 0.05M, $\Delta\Delta G_{25\,\text{Å}}$ becomes ~7.2$k_BT$ for $Z = -16e$ while still has a larger value of ~9.1$k_BT$ for $Z = -24e$. When 1:1 salt is very high (~0.5M), $\Delta\Delta G_{25\,\text{Å}}$ becomes ~0 for different $Z$'s. Such relation between the many-body effect and charges $|Z|$ on nanoparticles is coupled to ion-binding to nanoparticles. As shown in Fig. 3d–f, although the increase of number of the nanoparticles with lower charges can strengthen local electric field (between them) more strongly and can cause stronger increase in ion-binding fraction, the increase in total number of binding ions for the nanoparticles with lower charges is still slightly smaller than that for the nanoparticles with higher charges. For example, at 0.005M 1:1 salt and separation $x = 25$ Å, the number of binding ions per nanoparticle within 8 Å from surface of nanoparticles can increase by ~2.5 for $|Z| = 16e$ when the number of nanoparticles increases from 2 to 4. Such increase becomes a slightly larger value of ~2.8 for higher charge $|Z|$ ($=24e$). Therefore, the increase in ion-binding number by more nanoparticles is more pronounced for the nanoparticles with higher charges, and thus causes more apparent many-body effect for the PMFs. As a result, the global many-body effect would become stronger for more highly charged nanoparticles.

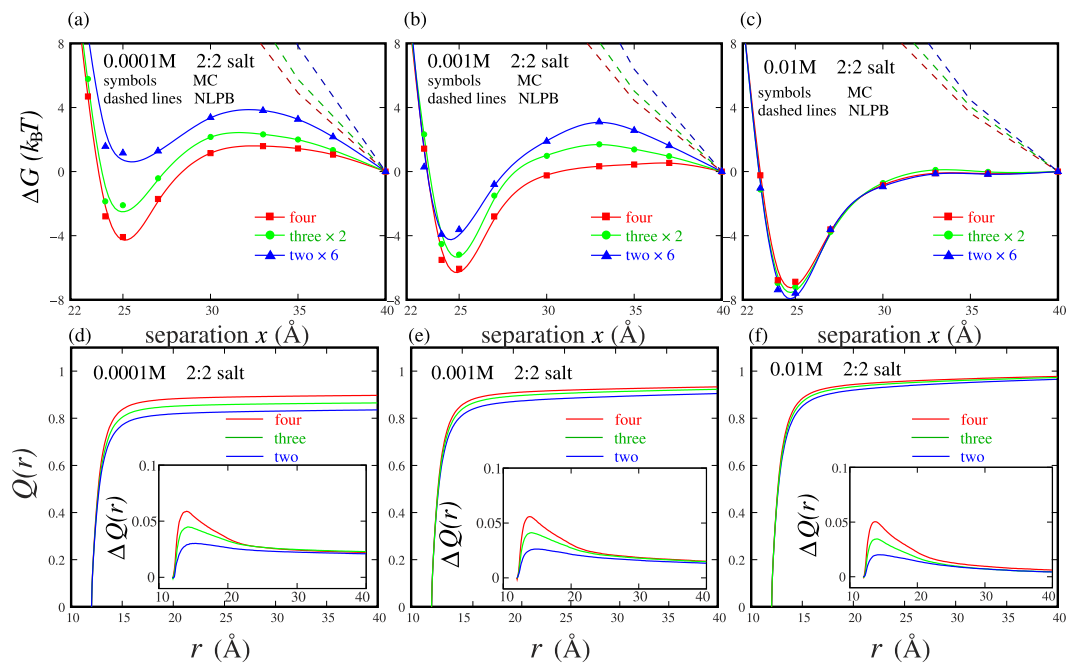

**Figure 4.** (**a**–**c**) The potentials of mean force $\Delta G_x$ as functions of the separation $x$ between nanoparticles with $-24e$ for the four-body systems in 2:2 salt solutions which are calculated respectively by the pair-wise potential of mean force abstracted from two-body, three-body and four-body systems (denoted respectively by two nanoparticles $\times 6$, three nanoparticles $\times 2$, and four nanoparticles). (**a**) 0.0001M, (**b**) 0.001M, and (**c**) 0.01M. The predictions from the nonlinear PB are shown as comparisons (dashed lines). (**d**–**f**) Net charge distribution $Q(r)$ per unit charge on nanoparticles as a function of distance $r$ around the nanoparticles with $x = 25$ Å in 0.0001M (**d**), 0.001M (**e**) and 0.01M (**f**) 2:2 salt solutions. The insets show the increase of $Q(r)$ due to the approaching of nanoparticles from $x = 40$ Å to $x = 25$ Å for the systems of two nanoparticles, three nanoparticles and four nanoparticles.

**Many-body PMFs in 2:2 salt solutions.** It have been previously shown that high concentration of divalent ions could induce aggregation of polyelectrolytes such as filamentous actins and filamentous virus fd[12,13] and could mediate an effective attraction between two like-charged nanoparticles[5,6,23]. Due to the important role of divalent ions (e.g., in RNA structure and function[2,4,9,27,52]), a special attention would be paid to the many-body effect in the PMF between nanoparticles in 2:2 salt solutions.

*Many-body effect "enhances" attractive PMF at low 2:2 salt.* As shown in Fig. 4a, at 0.0001M 2:2 salt, the pair PMF between two nanoparticles appears weakly repulsive, while for three-body and four-body nanoparticles, the PMFs become visibly attractive. For four-body nanoparticles, the additive assumption would overestimate the minimum PMF ($x \sim 25$ Å) by $\sim 5.2 k_B T$ at 0.0001M 2:2 salt; see Fig. 4a. This indicates that many-body effect would apparently enhance the attractive PMF at low 2:2 salt. As 2:2 bulk concentration is increased, pair PMF between two-body nanoparticles becomes attractive, and the enhancement on the attractive PMF by more nanoparticles would become weaker. For four-body nanoparticles, the additive assumption would overestimate the minimum PMF by $\sim 2.4 k_B T$ at 0.001M 2:2 salt; see Fig. 4b.

Such many-body effect on enhancing attractive PMF at low 2:2 salt is coupled to divalent ion-binding to nanoparticles. Due to the 2+ ionic charge, divalent ions interact with nanoparticles more strongly (than monovalent ions), and divalent ion-bridging between two nanoparticles may cause an effective attraction between them[5,6,13,23,26,42,43]; see Fig. 5a. As shown in Fig. 4d, at 0.0001M 2:2 salt, two nanoparticles with $x = 25$ Å can get $\sim 82\%$ ionic neutralization by 2+ ions within 8 Å from nanoparticle surface. Due to the low ion concentration ($\sim 0.0001$M) and the resultant high entropy penalty, such moderate ion-binding and ion-bridging between nanoparticles could only cause a weak repulsion between two nanoparticles[5,6,13,23]. However, with the increase of number of nanoparticles from 2 to 4, the ion-binding fraction increases from $\sim 82\%$ to $\sim 88\%$ within 8 Å from nanoparticle surface, which comes from the enhancement of electric field by more nanoparticles. More binding divalent ions would coordinate more strongly with adjacent nanoparticles and would consequently cause a visibly attractive PMF for four-body system. As 2:2 salt concentration is increased (e.g., to $\sim 0.001$M), more binding divalent ions would bridge nanoparticles more strongly due to the lowered ion-binding entropy penalty, and the pair PMF between two nanoparticles would become attractive. Simultaneously, the enhancement in ion-binding by more nanoparticles would become less pronounced due to the prior strong ion-binding for two-body system, which would cause weaker many-body effect on enhancing attractive PMF at higher 2:2 salt.

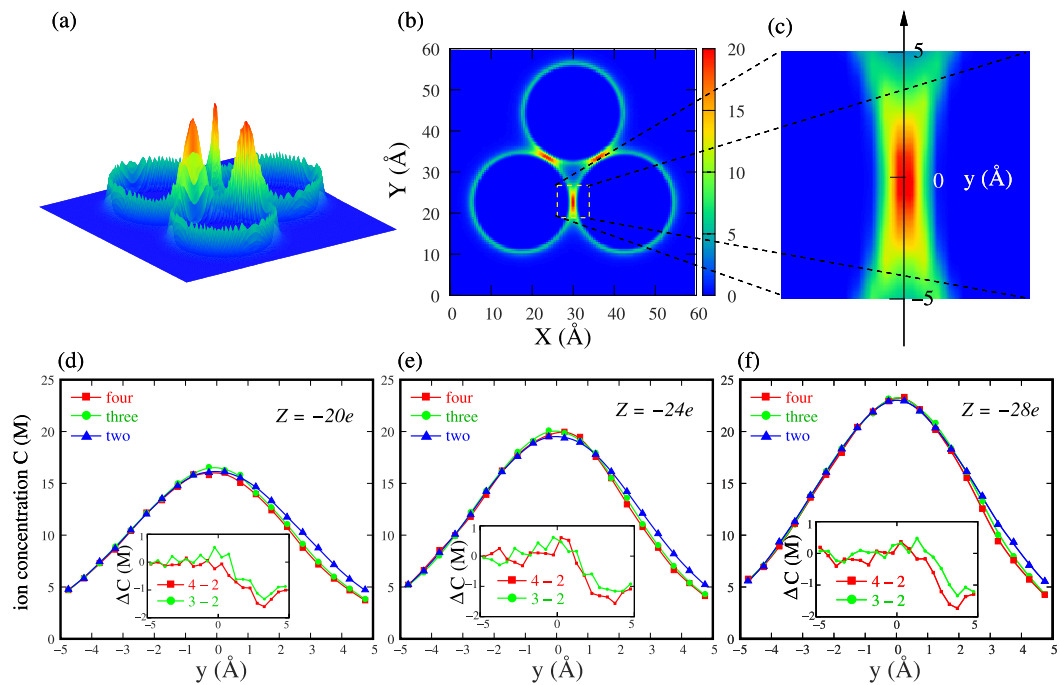

**Figure 5.** (**a**–**c**) The ion concentration distribution around three-body nanoparticles with $x = 25$ Å in 0.01 2:2 salt solution. The landscape plot (**a**) and density plot (**b**) show the ordered structure of ion bridge between like-charged nanoparticles. (**c**) The "ion bridge" zone has been highlighted and the coordinate $y$ has been built in order to indicate the detailed distribution of "ion bridge". (**d**–**f**) The detailed ion concentration distribution of "ion bridge" between nanoparticles with $x = 25$ Å in 0.01M 2:2 salt solutions for two-body, three-body and four-body nanoparticles with $Z = -20e$ (**d**), $Z = -24e$ (**e**) and $Z = -28e$ (**f**). The insets show the concentration differences between four-body and two-body systems (red; denoted by $4 - 2$) as well as those between three-body and two-body systems (green; denoted by $3 - 2$).

*Many-body effect "weakens" attractive PMF at high 2:2 salt.* When 2:2 salt becomes higher (e.g., ~0.01M), the pair PMF between two nanoparticles appears more attractive[5,6,23], while the many-body PMF is slightly less attractive than that calculated from the additive assumption with the pair PMF between two nanoparticles. As shown in Fig. 4c, the additive assumption overestimates the negative minimum of PMF by $\sim -0.7 k_B T$ for four-body nanoparticles at 0.01M 2:2 salt. Such many-body effect of attraction "weakening" at high 2:2 salt is in contrast to that at low 2:2 salt, and seems to be somewhat surprising, although it does not appear very strong. What causes such surprising many-body effect at high 2:2 salt?

To understand the elusive many-body effect at high 2:2 salt, we first examine the ion-binding profiles $Q(r)$ and find that two-body, three-body and four-body nanoparticles with $x = 25$ Å can all get very strong ionic neutralization of >90% within 8 Å from nanoparticle surface. The increase of nanoparticle number only causes very negligible increase in ion-binding fraction (<3%); see Fig. 4f. Such (negligible) increase in ion-binding number by more nanoparticles should not be responsible for the "weakened" many-body PMF at high 2:2 salt, since the enhancement of ion-binding would generally cause the enhancement of effective attraction. Then what causes such many-body effect of attraction weakening at high 2:2 salt? Since previous studies have shown that the highly ordered structure (ion bridge) of binding ions between two adjacent nanoparticles can induce an apparently effective pair attraction between them[5,6,11,12,23,32], we would like to pay attention to and analyze the structure of binding ions between two nanoparticles for two-body, three-body and four-body systems. Figure 5a illustrates the 2+ ion concentration distribution around nanoparticles. As expected, there are strongly binding ions of high order in the regions between nanoparticles (denoted by the very high concentration of ions), and such ordered structure of strong binding ions is responsible for the multivalent ion-mediated effective attraction between two like-charged particles[5,6,23,32]. Since the ion-binding in the region between nanoparticles is most important for the PMF, we would like to make a more detailed analysis on ion distribution in the region between two nanoparticles for two-body, three-body and four-body systems.

As shown in Fig. 5, when two nanoparticles interact strongly with each other at a short separation ($x = 24$ Å to 25 Å), the ion-binding structure between two-body nanoparticles appears most ordered and symmetrical, which would favor a strongest attractive PMF for the two nanoparticles. When more nanoparticles are added, the ion-binding would become less ordered for each pair of two nanoparticles in the many-body systems. Specifically, in the direction of nanoparticles added, binding ion concentration would decrease, which comes from the electrostatic repulsion from the new binding ions brought by the added nanoparticles. Such decrease can become as large as ~1.6M and ~4.5M when the two-body system is changed into four-body one at $x = 25$ Å and 24 Å, respectively; see Fig. 5d–f and Figure S3a–c in SI. Such less ordered ion-binding structure between two nanoparticles induced by more nanoparticles would not be most favorable for the two nanoparticles in causing an effective attraction between them. Therefore, the many-body effect would slightly "weaken" the attractive PMF at high 2:2 salt, as compared with the PMF from the calculations with the additive assumption.

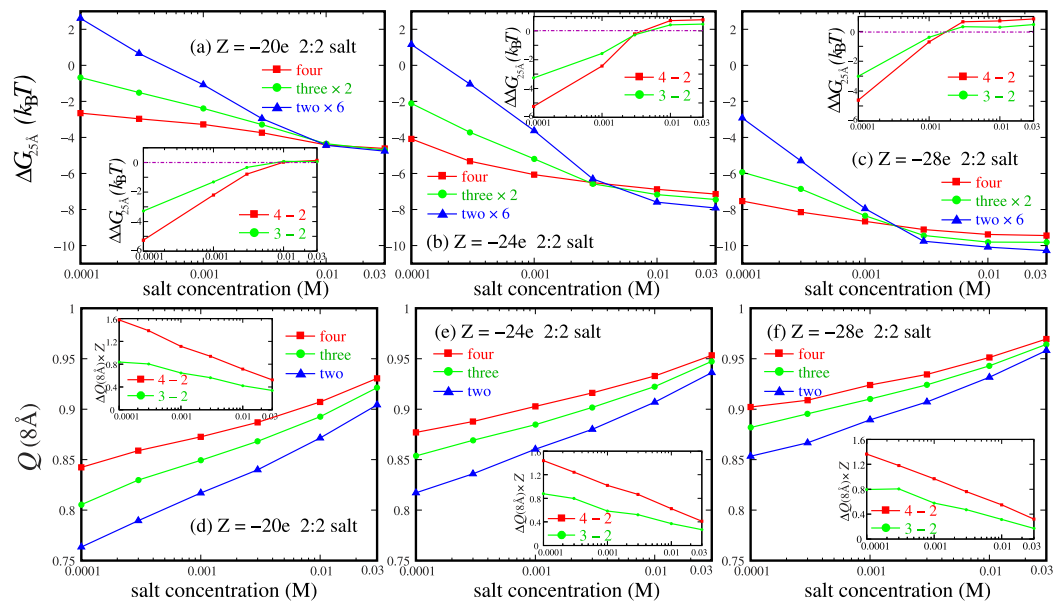

**Figure 6.** (**a–c**) $\Delta G_{25\,\text{Å}}\,(=G_{25\,\text{Å}} - G_{40\,\text{Å}})$ as functions of 2:2 salt concentration for the four-body systems which are calculated respectively by the pair-wise potential of mean force abstracted from two-body, three-body and four-body systems (denoted respectively by two nanoparticles ×6, three nanoparticles ×2, and four nanoparticles). (**a**) $Z = -20e$, (**b**) $Z = -24e$ and (**c**) $Z = -28e$ per nanoparticles. The insets show the PMF differences $\Delta\Delta G_{25\,\text{Å}}$ defined by $\Delta G_{25\,\text{Å}}(\text{four}) - \Delta G_{25\,\text{Å}}(\text{two}) \times 6$ (red; denoted by 4 − 2) and $\Delta G_{25\,\text{Å}}(\text{three}) \times 2 - \Delta G_{25\,\text{Å}}(\text{two}) \times 6$ (green; denoted by 3 − 2). (**d–f**) Net charge fraction within 8 Å from the surface of nanoparticles with $x = 25$ Å for $Z = -20e$ (**d**), $Z = -24e$ (**e**) and $Z = -28e$ (**f**). The insets show the difference in number of binding ions within 8 Å from the surface of nanoparticles between four- and two-body systems (red; denoted by 4 − 2) as well as that between three- and two-body systems (green; denoted by 3 − 2).

As analyzed above, the attractive PMF "weakening" by many-body effect at high 2:2 salt is proposed to come from the disordering of ordered ion-binding (bridging) structure caused by the electric field of new binding ions due to more nanoparticles. Therefore, such many-body effect is expected to become more pronounced for more strongly correlated system where ions can mediate strong effective pair attraction, e.g., for higher charges on nanoparticles.

*Many-body effect versus charge density on nanoparticles in 2:2 salt.* In analogy to 1:1 salt, we have also examined how the charge density on nanoparticles affects the many-body effect on PMF between nanoparticles in 2:2 salt solutions, and we use different charges of $Z = -20e$ and $-28e$ on each nanoparticle to calculate the PMFs for two-body, three-body and four-body systems. Figure 4 and S4 in SI show that, the many-body effect on PMFs is similar for nanoparticles with different $Z$'s in 2:2 salt solutions and $\Delta G_{25\,\text{Å}}\,(=G_{25\,\text{Å}} - G_{40\,\text{Å}})$ can approximately describe the strength of attractive PMFs between nanoparticles in 2:2 salt solutions. As shown in Fig. 6a–c, and also in Fig. 4 and Figure S4, the many-body effect has analogous salt-concentration dependence for nanoparticles with different $Z$'s: at low (~0.0001M) 2:2 salt, the many-body effect would enhance the attractive PMF, while it would slightly "weaken" the attractive PMF at high (~0.01M) 2:2 salt, compared with the PMFs from the calculations with the additive assumption.

Despite of the similarity for different $Z$'s, there are also several different features for different $Z$'s. First, as $|Z|$ is increased from 20e to 28e, the pair PMF between two-body nanoparticles at low 2:2 salt could transit from an (weakly) effective repulsion to an effective attraction, which is attributed to the stronger ion-binding and ion-bridging effect between two nanoparticles with higher $|Z|$[23]. Second, for higher $|Z|$ at high 2:2 salt (e.g., 0.01M), the many-body effect of attraction "weakening" becomes slightly stronger. This is because the involvement of more nanoparticles with higher $|Z|$ would cause (slightly) stronger disordering of order structure of binding ions between a pair of nanoparticles, as discussed above and also shown in Fig. 5 and Figure S3 in SI. Third, the crossover salt concentration $c_c$ from the attraction enhancement at low 2:2 salt to the attraction "weakening" at high 2:2 salt by many-body effect decreases with the increase of $|Z|$. At $c_c$, the PMFs between the nanoparticles in 2:2 salt solutions are exactly additive. This is reasonable because nanoparticles with higher $|Z|$ can get strong ionic neutralization even at lower 2:2 salt as shown in Fig. 6d–f, and the involvement of more nanoparticles would cause the disordering of ordered structure of binding ions between two nanoparticles. Consequently, the attraction "weakening" caused by many-body effect can occur at lower 2:2 salt for higher $|Z|$. Actually, $c_c$ is the balance point for the competition between two contributions to the many-body effect: the attraction "enhancement" at low 2:2 concentration since more nanoparticles can get apparently stronger ion-binding, and the attraction "weakening" at high 2:2 concentration since more nanoparticles will cause the disordering of the ordered structure of "bridging" ions between two nanoparticles.

In addition, for nanoparticles in 2:2 salt solutions, the predictions from the nonlinear PB are shown for comparisons. Figure 4 shows that the nonlinear PB always predicts the effective repulsion between nanoparticles. Such significant discrepancy is attributed to the neglect of ion-ion correlations in the nonlinear PB[11,47]. The nonlinear PB is based on mean-field approximation, where ions are approximated as continuous fluid-like particles moving independently in a mean electric field[38–40,47]. Thus, the discrete properties of ions such as ion size and ion-ion correlations are ignored in the PB theory. However, such discrete properties of ions are accounted for in the MC simulations; shown in Model and Method. The electric field from nanoparticles and inter-ion correlation can drive the binding ions to organize to a correlated low-energy state, which can cause an effective attraction between nanoparticles[11,23,24], and is beyond the description of the PB theory[5,6,11,24].

### Comparisons with previous studies.
In the present work, the many-body effect in PMF between like-charged nanoparticles has been calculated and analyzed extensively with the MC simulation and nonlinear PB over the wide ranges of monovalent and divalent salts. For monovalent salt, the PMFs between nanoparticles at low salt concentration are strongly repulsive and the many-body effect could apparently weaken such repulsive PMFs. With the increase of salt concentration, the many-body effect would become weaker, and approximately convergent to zero at very high salt. Wu *et al.*[19], Ikeda *et al.*[35] and Kreer *et al.*[36] have respectively obtained a short-ranged attractive triplet force for different like-charged spheres at low monovalent salt with computer simulations. Merrill *et al.*[18] and von Grünberg *et al.*[20,21] have experimentally measured the triplet force respectively, and found that the triplet force is attractive at low salt and would become less attractive for higher salt. Our predictions are in good agreement with the above related simulational and experimental studies[18–21,35,36]. Besides, a recent coarse-grained molecular dynamics simulation study for three-body charged dendrimers has shown that the many-body effect would increase the short-ranged repulsion between dendrimers[34]. This is not contradictory with the above results since the flexible star-like dendrimers can accommodate counterions inside and are very different from the hard-core-like nanoparticles[34].

For divalent salt, the two-body PMFs between nanoparticles at low salt concentration are weakly repulsive and the many-body effect could induce an attractive PMF between nanoparticles. However, at higher divalent salt, the two-body PMF is apparently attractive while the many-body effect would slightly weaken such attractive PMF. Wu *et al.* have found that the three-body effect is weak for nanoparticles with slightly low charge density at a high divalent concentration[19]. Tan and Chen have employed the TBI theory for three parallel DNAs in divalent ion solutions, which shows that the three-body effect would transit a weakly repulsive PMF between DNAs to a weakly attractive PMF at low divalent salt while would slightly weaken the strong attractive PMF between DNAs at high divalent salt[24]. Our predictions also agree with these previous studies for divalent salt[19,24]. Furthermore, in the present work, the detailed ion-binding properties have been analyzed to reveal the mechanism for the many-body effect in PMF between like-charged nanoparticles over the wide ranges of monovalent and divalent ion concentrations.

### Conclusions
In this work, we have employed the MC simulation with the thermodynamics-integration and the nonlinear PB theory to investigate the many-body effect in the potentials of mean force between highly like-charged nanoparticles for two-body, three-body and four-body systems, and the study covers the wide ranges of ionic conditions including 1:1 and 2:2 salts with different concentrations. The additivity as well as non-additivity of many-body potentials of mean force have been examined and analyzed, especially in multivalent salt solutions. Through the systematic calculations, we have obtained the following conclusions:

1. At high 1:1 salt, the PMF between like-charged nanoparticles is repulsive and additive, while at low 1:1 salt, the additive assumption would strongly overestimate the repulsive many-body PMF.
2. At low 2:2 salt, the pair PMF between two like-charged nanoparticles is weakly repulsive (or weakly attractive depending on nanoparticle charge) and many-body effect could cause an apparently attractive PMF. At high 2:2 salt, the pair PMF is apparently attractive while many-body effect could cause a slightly "weakened" attractive PMF compared with that from the additive assumption.
3. The many-body effect at low 1:1/2:2 salts comes from the fact that more nanoparticles would lead to significantly stronger ion-binding which could cause stronger ionic neutralization for 1:1 salt and stronger divalent ion-bridging effect between nanoparticle pairs for 2:2 salt. The many-body effect at high 2:2 salts is attributed to that more nanoparticles would lead to the disordering of the ordered ion-bridge structure between nanoparticles for two-body system.
4. For more highly charged nanoparticles, the many-body effect could become more apparent: (*i*) at low 1:1 salt, the additive assumption would overestimate the repulsive many-body PMF more significantly; and (*ii*) at high 2:2 salt, the attraction "weakening" induced by many-body effect would become more apparent.

In addition, for nanoparticles in 1:1 and 2:2 salt solutions, the predictions from the nonlinear PB are shown for comparisons. The above conclusions have been examined for highly like-charged particles with different charges over extensive ionic conditions. Therefore, the present work would provide an overall picture and direct illustration on the many-body effect in the PMF between highly charged nanoparticles. The many-body effect on PMF between like-charged nanoparticles in multivalent salt has been seldom covered previously and has been shown here to be somewhat surprising and interesting. Our predictions for 2:2 salt are also helpful for understanding the driving force for the aggregation of polyelectrolytes such as F-actins and DNAs in multivalent salt solutions which have been investigated by various experiments[10,13,25,26,31,42,49]. The many-body effect of PMF indicates that the additivity assumption based on two-body PMF would overestimate the repulsive force between multiple polyelectrolytes at low 2:2 salt and consequently would overestimate the critical salt concentration for the aggregation of polyelectrolytes[13,42,43]. The many-body effect of PMF at high 2:2 salt solutions suggests that the additivity assumption would slightly overestimate the attractive force between polyelectrolytes in aggregates at high multivalent salt[10,13,25,26,31,42,49], which would slightly overestimate the stability of polyelectrolyte aggregates and would also contribute to the deviation between experimental measurements and theoretical predictions from additivity assumption on osmotic pressure for DNA aggregates[11,32]. Certainly, the direct and quantitative experimental measurements on many-body effect for multivalent salt are still highly desirable, which may motivate the development of novel applications such as self-assembly of nanoparticle crystal in multivalent salt solution.

However, the present work on many-body effect is limited to the number of 4 for like-charged nanoparticles. The involvement of more nanoparticles than 4 would bring next-nearest-neighbor PMF between nanoparticles[18] and consequently bring more complexity in analyzing the mechanism in many-body effect. Although at high 2:2 salt concentration, many-body effect causes the weakening of the attractive PMF, such weakening is not very significant and the global attractive PMF between like-charged nanoparticles is still strong in spite of the many-body effect. Since the disordering caused by more nanoparticles should be dominated by the nearest-neighbor nanoparticles, the weakening of attractive PMF is expected to become saturated after the nearest-neighbor nanoparticles are fulfilled. Afterwards, the like-charged nanoparticles would still have tendency for aggregation at high 2:2 salt and the size of aggregates would depend on the strength of attractive PMF and the translational entropy of nanoparticles[53].

The present work also involves some simplifications and approximations. First, the solvent water molecules are modeled as a continuum medium with a high dielectric constant and the dielectric discontinuity at the interface between nanoparticles and solvent is ignored. Such ignorance on dielectric discontinuity might only slightly affect the ion-binding near nanoparticle surface[54] since nanoparticles may have much higher dielectric constant than vacuum due to the solvent invasion[55]. Furthermore, the enhancements of (repulsive) ion self-energy and (attractive) ion-nanoparticle could be partially cancelled out[47]. Second, for simplicity, the hard-core potential is used instead of softer Lennard-Jones potential, which would only have slight effect on the ion-binding near nanoparticle surface and the potential of mean force. Finally, the separation of 40 Å is used as the outer-reference distance, which may not be far enough for some ionic conditions. Nevertheless, we have presented an overall picture on the many-body effect for the ion-mediated potential of mean force for like-charged particles, which would be very helpful for understanding the complicated many-body charged systems.

## Model and Method

**Model systems for like-charged nanoparticles.**   In this work, for generality and simplicity, we use two, three and four charged macro-spheres as models for nanoparticles to study the many-body effect in PMFs between like-charged nanoparticles. As shown in Fig. 1, the spherical nanoparticles are set up with high spatial symmetry: (*i*) two-body nanoparticles are in a line and there is only one pairwise interaction between them; (*ii*) three-body nanoparticles are set up as a equilateral triangle and there are three pairwise interactions between them; (*iii*) four-body nanoparticles are set up as a regular tetrahedron and there are six pairwise interactions between the four nanoparticles. These configurations with high spatial symmetry are based on the previous related studies[19–21,34–36], and would be convenient for the detailed analysis on the many-body effect in PMF and the related ion distributions[5,6,19,23]. The (two, three or four) nanoparticles are immersed in 1:1 or 2:2 salt solutions where ions are simplified as small spheres with the equal radii and different ionic charges. Counterions are added to keep charge neutralization of the system and are approximated as the cations with the same valence and radii as the salt cations[5,6,19,22,23]. During the simulations, both counterions and cations are treated as identical ions[5,6,19,22,23]. Also, water is implicitly modeled as continuum medium with dielectric constant $\varepsilon = 78$[5,6,23]. The radii of ions and nanoparticles are taken as 2 Å and 10 Å respectively, which are consistent with the previous related studies[5,6,19,23] and the radii of nanoparticles are approximately equal to the (radial) radii of B-DNA and A-RNA[1,3,11,24,32]. The charges $Z$ of the nanoparticles are taken as $-24e$, and the surface charge density is in the range of the distribution of RNA surface charge density[56]. For studying charge density effect, we have also made the additional calculations for other charges on nanoparticles ($Z = -16e$ and $-20e$ for 1:1 salt solutions, and also $Z = -20e$ and $-28e$ for 2:2 salt solutions).

**Calculating PMFs with Monte Carlo simulation.**   The interactions of the systems are simplified into two parts: the electrostatic interaction $U_{el}$, and the excluded-volume interaction $U_{ex}$. The $U_{el}$ between particle *i* and particle *j* can be written as

$$U_{el} = \frac{q_i q_j}{4\pi\varepsilon_0 \varepsilon r_{ij}}, \tag{2}$$

where $q_i$ and $q_j$ are the charges on particles (ions or nanoparticles) *i* and *j*, and $r_{ij}$ is the center-to-center distance between them. $\varepsilon_0$ is the permittivity of vacuum and $\varepsilon$ is the dielectric constant of water. The $U_{ex}$ between particles *i* and *j* is approximately given by the hard-core potential

$$U_{ex} = \begin{cases} \infty & \text{for } r_{ij} < \sigma_{ij}; \\ 0 & \text{for } r_{ij} \geq \sigma_{ij}, \end{cases} \tag{3}$$

where $\sigma_{ij}$ is the sum of radii of particles $i$ and $j$. The hard-core potential is involved to avoid overlap of two particles (both ions and nanoparticles).

*Thermodynamics-integration method for calculating PMFs.* For the above described system of charged nanoparticles and mobile ions, the configurational partition function $Z_U$ can be written as

$$Z_U = \int e^{-\beta U} (d\mathbf{r})^{3N}, \qquad (4)$$

where $U$ is the interaction energy of the whole system. $\beta = 1/k_BT$, where $T$ is absolute temperature in Kelvin and $k_B$ is the Boltzmann constant. $N$ denotes the total number of ions in the system.

If the configurational partition function is made as a function of a control variable $\lambda \in [0, 1]$, the free energy $G$ will be a function of $\lambda$ as well,

$$G(\lambda) = -k_BT \ln Z_U(\lambda). \qquad (5)$$

According to Eqs (4) and (5), the differentiation of $G(\lambda)$ with respect to $\lambda$ gives

$$\frac{\partial G(\lambda)}{\partial \lambda} = k_BT \left\langle \frac{\partial (\beta U)}{\partial \lambda} \right\rangle, \qquad (6)$$

where $\langle \rangle$ implies the canonical ensemble average. The free energy difference between two states with $\lambda = 0$ and $\lambda = 1$ for the system is then given by

$$\Delta G = G_{\lambda=1} - G_{\lambda=0} = k_BT \int_0^1 \left\langle \frac{\partial (\beta U)}{\partial \lambda} \right\rangle d\lambda, \qquad (7)$$

which is the generalized form of the TI method[57]. In this work, we use $\beta$ as the control variable $\lambda$, and Eq. (7) leads to[53]

$$G_{\beta_T} = G_{\beta=0} + k_BT \int_0^{\beta_T} \langle U \rangle_\beta d\beta, \qquad (8)$$

where $\beta_T$ is the value of $\beta$ of the target temperature (T = 25 °C). $\langle U \rangle_\beta$ is the average interaction energy at $\beta$, and $G_{\beta=0}$ stands for the free energy when $\beta = 0$ (i.e., $T \to \infty$). In practice, to deal with the integration in Eq. (8), we divide $[0, \beta_T]$ into 20 equal intervals and use the summation instead of integration. The Monte Carlo with temperature annealing has been employed for each specific $\beta$ which is changed from 0 to $\beta_T$ with the temperature annealing process[58], in order to obtain $\langle U \rangle_\beta$ more smoothly and efficiently. Therefore, the PMF $\Delta G_x$ can be calculated as

$$\Delta G_x = G_x - G_{x_{ref}}, \qquad (9)$$

where $G_x$ is the free energy of the nanoparticles with the center-to-center separation $x$. $x_{ref}$ is the outer-reference distance, and for simplicity, $x_{ref}$ is taken as 40 Å for all ionic conditions in our calculations[23].

*Monte Carlo simulation.* The Metropolis algorithm has been employed to calculate the average interaction energy $\langle U \rangle_\beta$ for each $\beta$[59]; see Eq. (8). Each MC simulation starts from an initial configuration with (two, three or four) fixed nanoparticles and randomly distributed ions at $\beta = 0$. Every random trial move of an ion in the simulation cell would yield a new configuration. The energy change $\Delta U$ due to the trial move can be calculated, and the new configuration is accepted with the probability $P = \min(\exp(-\Delta U/k_BT), 1)$. Repeat the process until the system reaches an equilibrium, and then the average energy of the system can be calculated at a $\beta$. The MC with temperature annealing is used and the last configuration of ions in equilibrium is taken as the initial state for the following MC simulation at a higher $\beta$ (lower temperature)[58]. The simulation cells are cuboids. To diminish the boundary effect, the cell size is generally kept at least eight times larger than the Debye-Hückel length for ionic solutions[23], and the periodic boundary condition has been also employed. Practically, at each $\beta$, the MC process is continued until the difference in mean energy $\Delta \langle U \rangle_\beta (=|\langle U \rangle_\beta(t) - \langle U \rangle_\beta(t-10^4)|)$ converges at least to $0.005k_BT$ for 1:1 salt and $0.001k_BT$ for 2:2 salt, where $t$ is the total MC steps in equilibrium and $10^4$ is the span of MC steps for examining convergence.

**The nonlinear Poisson-Boltzmann theory and numerical solution.** The nonlinear PB is a well-established mean-field theory, which has been used to solve electrostatic problems of macromolecules for many years[40,60]. For solutions with polyelectrolyte and ions, the electrostatic potential $\psi(\mathbf{r})$ is given by the Poisson equation[38,47]

$$\nabla \cdot [\varepsilon_0 \varepsilon(\mathbf{r}) \nabla \psi(\mathbf{r})] = -4\pi \{\rho_f + \sum_\alpha z_\alpha e c_\alpha(\mathbf{r})\}, \qquad (10)$$

where $\rho_f$ is the charge density of fixed charges. $z_\alpha e$ is the charge of ion species $\alpha$. $\varepsilon_0$ is the permittivity of vacuum and $\varepsilon(\mathbf{r})$ is the permittivity in continuous medium. $\psi(\mathbf{r})$ can be solved by assuming that the concentration $c_\alpha(\mathbf{r})$ of ion species $\alpha$ obeys the Boltzmann distribution

$$c_\alpha(\mathbf{r}) = c_\alpha^0 e^{-\beta z_\alpha e \psi(\mathbf{r})}, \tag{11}$$

where $c_\alpha^0$ is the bulk concentration of ion species $\alpha$.

With $\psi(\mathbf{r})$ and $c_\alpha(\mathbf{r})$ from the nonlinear PB, the electrostatic free energy for the system can be calculated by[47,52,60]

$$\begin{aligned} G_{NLPB} = \ & U_M + \frac{1}{2}\int\sum_\alpha c_\alpha(\mathbf{r})z_\alpha e[\psi(\mathbf{r}) + \psi'(\mathbf{r})]\mathrm{d}^3\mathbf{r} \\ & + k_B T \int\sum_\alpha \{c_\alpha(\mathbf{r})\ln\frac{c_\alpha(\mathbf{r})}{c_\alpha^0} - c_\alpha(\mathbf{r}) + c_\alpha^0\}\mathrm{d}^3\mathbf{r}, \end{aligned} \tag{12}$$

where $U_M$ is the Coulombic energy for $M$ nanoparticles in the form of Eq. (2). $\psi'(\mathbf{r})$ is the electrostatic potential without diffusive salt ions. Afterwards, the PMF from the nonlinear PB can be computed through Eq. (9).

In the work, we use the three-dimensional algorithm developed in the TBI theory to numerically solve the nonlinear PB equation[47,52]. A thin layer of one ion radius is added to nanoparticle surface to account for the excluded volume layer of ions, and the three-step focusing process has been used to compute the detailed electrostatic potential near nanoparticle surface[47,52]. The grid size in the first run of the three-step focusing process depends on the salt concentration used. Generally, we choose a grid-size six times larger than the Debye length from nanoparticle surface, to effectively include the salt effect in solution. The resolution of the first run varies with the grid size to make the iterative process computationally efficient. The grid size $(L_x, L_y, L_z)$ in the second and the third runs are kept at (125 Å, 120 Å, 120 Å) and (75 Å, 70 Å, 70 Å), respectively, and the corresponding resolutions are 0.5 Å per grid and 0.25 Å per grid, respectively. As a result, the number of the grid points is $251 \times 241 \times 241$ in the second and $301 \times 281 \times 281$ in the third run. The iteration for a run is continued until the electrostatic potential change $\delta\psi(\mathbf{r})$ for an iteration is less than $10^{-4}\,k_B T/e$[47,52].

## Acknowledgements

We are grateful to Profs. Shi-Jie Chen (Univ. Missouri), Jianzhong Wu (Univ. California), Wenbing Zhang (Wuhan Univ.), and Dr. Yuan-Yan Wu (Yangzhou Univ.) for valuable discussions. This work is supported by the National Key Scientific Program (973)-Nanoscience and Nanotechnology (No. 2011CB933600), the National Science Foundation of China grants (11175132, 11074191, 11374234 and 11547310), and the Program for New Century Excellent Talents (Grant No. NCET 08-0408).

## Author Contributions

Z.J.T. and X.Z. designed the research; X.Z., J.S.Z. and Y.Z.S. performed the calculations; Z.J.T., X.Z., J.S.Z. and X.L.Z. analyzed the data; X.Z. and Z.J.T. wrote the manuscript; All authors discussed the results and reviewed the manuscript.

## Additional Information

**Supplementary information** accompanies this paper at http://www.nature.com/srep

**Competing financial interests:** The authors declare no competing financial interests.

**How to cite this article**: Zhang, X. *et al.* Potential of mean force between like-charged nanoparticles: Many-body effect. *Sci. Rep.* **6**, 23434; doi: 10.1038/srep23434 (2016).

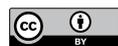